\documentclass
 [aps,prl,reprint,preprintnumbers,showpacs,floatfix,nofootinbib,superscript
 address,longbibliography]{revtex4-2}
\usepackage[utf8]{inputenc}
\usepackage[T1]{fontenc}
\usepackage{amsmath,amsfonts,amssymb}
\usepackage{epsf,amsmath,bbold,amsfonts,stmaryrd}
\usepackage{hhline}
\usepackage{mathrsfs}
\usepackage{mathtools}
\usepackage[dvipsnames]{xcolor}
\usepackage{multirow,tabularx}
\usepackage{graphicx}
\usepackage{natbib}

\usepackage{xspace}
\usepackage{xstring}
\usepackage{titlesec}
\usepackage{parskip}
\usepackage{calc}
\usepackage{bm}

\allowdisplaybreaks
\usepackage{braket}

\newrobustcmd{\pea}[1]{%
 	\emph{#1}\textbf{\ \ \ ---}
 }
\titleformat{\paragraph}[runin]{\normalfont\normalsize\bfseries}
 {\emph\theparagraph}{1em}{\pea}

\newcommand*{\ie}{i.e.\@\xspace}
\newcommand*{\eg}{e.g.\@\xspace}
\newcommand*{\cf}{c.f.\@\xspace}
\newcommand*{\fig}{fig.\@\xspace}
\newcommand*{\eq}{eq.\@\xspace}
\newcommand*{\eqs}{eqs.\@\xspace}

\def\f{\frac}

\def\G{\Gamma}

\def\mc{\mathcal}

\def\m{\mu}
\def\n{\nu}

\def\p{\partial}
\def\vp{\varphi}

\def\s{\sigma}

\def\vp{\varphi}

\def\z{\zeta}

\def\be{\begin{equation}}
\def\ee{\end{equation}}

\def\bea{\begin{eqnarray}}
\def\eea{\end{eqnarray}}

\def\ba{\begin{array}}
\def\ea{\end{array}}

\def\bc{\begin{center}}
\def\ec{\end{center}}

\def\bl{\begin{flushleft}}
\def\el{\end{flushleft}}

\def\br{\begin{flushright}}
\def\er{\end{flushright}}

\def\bi{\begin{itemize}}
\def\ei{\end{itemize}}

\def\bt{\begin{tabular}}
\def\et{\end{tabular}}

\def\be{\begin{equation}}
\def\ee{\end{equation}}

\def\bea{\begin{eqnarray}}
\def\eea{\end{eqnarray}}

\def\f{\frac}

\def\p{\partial}

\usepackage{hyperref}
\hypersetup{%
     pageanchor= false,%
     colorlinks = true,%
     allcolors= red}%

\usepackage{xifthen}
\newcommand*\diff{\mathrm{d}} 
\newcommand*\ldiff[2][]{ \ifthenelse{\isempty{#1}}{ \diff
#2}{\diff^#1#2} \,} 
\let\limitint\int 
\renewcommand{\int}{\limitint \!} 

\makeatletter 
    
\renewcommand\onecolumngrid{%
\do@columngrid{one}{\@ne}%
\def\set@footnotewidth{\onecolumngrid}%
\def\footnoterule{\kern-6pt\hrule width 1.5in\kern6pt}%
}%

\makeatother

\begin{document}

\title{On the Gravitational Origin of the QCD Axion}

\author{Georgios K. Karananas}
\email{georgios.karananas@physik.uni-muenchen.de}
\affiliation{Arnold Sommerfeld Center, Ludwig-Maximilians-Universit\"at,
 Theresienstraße 37, 80333 M\"unchen, Germany}
\author{Mikhail Shaposhnikov}
\email{mikhail.shaposhnikov@epfl.ch}
\affiliation{Institute of Physics, \'Ecole Polytechnique F\'ed\'erale de
 Lausanne, CH-1015 Lausanne, Switzerland}
\author{Sebastian Zell}
\email{sebastian.zell@lmu.de}
\affiliation{Arnold Sommerfeld Center, Ludwig-Maximilians-Universit\"at,
 Theresienstraße 37, 80333 M\"unchen, Germany}
\affiliation{Max-Planck-Institut für Physik, Boltzmannstr. 8, 85748 Garching
 b.\ M\"unchen, Germany}

\begin{abstract} 

Gravity can give rise to (pseudo)scalar fields, for instance due to torsion.
In particular, axions of gravitational origin have been proposed as a minimal
and compelling solution to the strong CP problem. In this work, we critically
examine the feasibility of this approach. We demonstrate that models in which
the scalar field couples to fermionic currents only through derivatives do
not yield a satisfactory axion. Moreover, we identify the necessary
conditions for generating a gravitational axion through quantum effects,
highlighting Weyl-invariant Einstein-Cartan gravity as a promising
theoretical setting.

\end{abstract}

\maketitle
\paragraph*{The strong CP problem}

At present, all known phenomena at microscopic scales can be accurately
described by the Standard Model of particle physics (SM). Having withstood
extensive experimental scrutiny, the SM was formally completed with the
discovery of the Higgs boson \cite{ATLAS:2012yve,CMS:2012qbp}. Arguably,
however, our understanding of the SM is still far from complete. Among the
most important open questions is the strong CP problem:  the puzzling
experimental fact that quantum chromodynamics (QCD), the theory of strong
interactions, preserves CP symmetry to an extraordinary accuracy, despite
having no apparent reason to do so. 

A compelling resolution to the strong CP problem relies on the introduction of
a new pseudoscalar particle -- the axion \cite
{Peccei:1977hh,Weinberg:1977ma,Wilczek:1977pj} -- which dynamically
suppresses CP violation in QCD (see reviews \cite
{Hook:2018dlk,DiLuzio:2020wdo}).  In addition to solving the strong CP
problem, the axion is a well-motivated dark matter candidate and thereby
addresses a crucial open issue in cosmology. Extensive experimental efforts
are underway to detect the axion \cite
{ALPS:2009des,0910.5914,Irastorza:2011gs,CAST:2011rjr,Budker:2013hfa,
Armengaud:2014gea,Kahn:2016aff,Brubaker:2016ktl,Caldwell:2016dcw,
DMRadio:2022pkf};~for an overview, see \cite{Adams:2022pbo}.

\paragraph*{Axion origins} 

In the oldest proposals, the axion arises from the spontaneous breaking of a
global Peccei-Quinn (PQ) symmetry \cite
{Peccei:1977hh,Weinberg:1977ma,Wilczek:1977pj}. However, the most popular
models \cite{Kim:1979if,Shifman:1979if,Zhitnitsky:1980tq,Dine:1981rt} require
the introduction of at least five degrees of freedom in addition to the
axion. Moreover, global symmetries are generally expected to be violated by
quantum gravitational effects (see \eg \cite{Kallosh:1995hi}), posing a
challenge to the theoretical consistency of this approach, as
Planck-suppressed operators could reintroduce CP violation \cite
{Kamionkowski:1992mf,Holman:1992us,Barr:1992qq}.

These concerns have motivated the development of alternative axion models,
including those based on extra dimensions \cite{Witten:1984dg}
(see review \cite{Reece:2024wrn}), or constructions where the axion is the
St\"uckelberg field of the QCD gauge redundancy \cite
{Dvali:2005an,Dvali:2013cpa,Dvali:2017mpy,Dvali:2022fdv}. Rather than being
an obstacle, gravity itself may play a constructive role: the axion could be
of purely gravitational origin \cite
{Mielke:2006zp,Mercuri:2009zi,Mercuri:2009zt,Mercuri:2009zt,Lattanzi:2009mg,
Castillo-Felisola:2015ema,Karananas:2018nrj,Karananas:2024xja}. This scenario
is especially appealing due to its minimality, requiring no ingredients
beyond the SM and General Relativity~(GR). Furthermore, recent work indicates
that the existence of an axion is a requirement for the consistency of
quantum gravity itself~\cite{Dvali:2018dce,Dvali:2022fdv,Dvali:2023llt}. 

\paragraph*{In this work} 

We critically examine the proposal that the axion originates from purely
gravitational dynamics. First, we show that the previously used arguments are
not successful. Our proof is almost trivial and relies on the fact that the
relevant operator (shown in \eq \eqref{fundamentalOperator}) as considered in
existing proposals, fails to break the shift symmetry and hence does not
generate any potential. If a scalar obeys an exact shift symmetry, it is
unable to dynamically solve the strong CP problem and hence the corresponding
particle cannot be an axion.

Second, we demonstrate that, generally, in Einstein-Cartan (EC) gravity, the
shift symmetry of the gravitational axion is explicitly broken.  We discuss
whether this gravitational breaking can induce the interactions of the axion
with quarks and gluons necessary for the solution of the strong CP problem. 

Third, we point out that the Weyl-invariant subclass~\cite
{Karananas:2024xja} of EC gravity is special. The requirement of exact Weyl
symmetry at the {\em quantum level} allows the control of radiative
corrections and leads to a tower of higher-dimensional operators which can
enable the gravitational scalar to assume the role of the QCD axion.
 
\paragraph*{Review of QCD axion}

At the heart of the strong CP problem lies the operator
\be 
\label{topologicalDensity}
\mathcal{L} \supset \frac{\alpha}{4\pi}\bar\theta 
\,\text{Tr}\, G_{\mu\nu} \tilde{G}^{\mu\nu} \;,
\ee
where $\alpha=g^2/(4\pi)$ with $g$ the QCD gauge coupling and $\bar\theta$ is
the physical CP violating parameter, \ie $\bar\theta$ is the sum of the
vacuum superselection angle of QCD and the CP violating phase of the quark
mass matrix. Moreover, $G_{\mu\nu}$ and $\tilde{G}_{\mu\nu}$ are the field
strength tensor of QCD and its dual, respectively, and the trace is taken
over color indices.

Although \eq \eqref{topologicalDensity} is a total derivative and hence does
not contribute perturbatively to equations of motion, it causes
non-perturbative CP violation \cite
{Callan:1976je,Jackiw:1976pf,Shifman:1979if}. In turn, this translates in the
neutron having an electric dipole moment $d_n \simeq 10^{-16}\bar\theta~
{\rm e~cm}$; when compared with the experimentally measured value~\cite
{Abel:2020pzs} $d_n^{\rm exp.} \lesssim 10^{-26}~{\rm e~cm}$, it dictates
that
\be
\bar\theta \lesssim 10^{-10} \;.
\ee
The strong CP problem corresponds to the question why $\bar\theta$ is so
small.

The key of the axion solution \cite
{Peccei:1977hh,Weinberg:1977ma,Wilczek:1977pj} is to effectively promote
$\bar\theta$ to a dynamical variable --- a canonically normalized
pseudoscalar field $a$ --- which extends \eq \eqref
{topologicalDensity} as
\be
\label{topologicalDensityAxion}  
\mathcal{L} \supset \frac{\alpha}{4\pi} \left(\bar\theta - \f{a}{f_a}\right) 
\text{Tr}\, G_{\mu\nu} \tilde{G}^{\mu\nu} \;,
\ee
where $f_a$ is the axionic decay constant. Since $\text{Tr}\, G_
{\mu\nu} \tilde{G}^{\mu\nu}$ is a total derivative, the coupling \eqref
{topologicalDensityAxion} leads to a perturbative shift symmetry for the
axion. Non-perturbative QCD effects break this shift symmetry and induce a
potential for the axion \cite
{Peccei:1977hh,Weinberg:1977ma,Wilczek:1977pj}, which can be well
approximated as \cite{DiVecchia:1980yfw,GrillidiCortona:2015jxo} (see also
discussion in \cite{DiLuzio:2020wdo}) 
\be
\label{ALP_pot}   
V = -m_\pi^2 f_\pi^2 \sqrt{1-\frac{4 m_u m_d}{(m_u+m_d)^2}
\sin^2\left[\frac{1}{2}\left(\bar\theta - \f{a}{f_a}\right)\right]} \ .
\ee
Here $m_\pi$ and $f_\pi$ are the pion mass and decay constant, respectively,
while $m_u$ and $m_d$ correspond to the masses of the up- and down- quarks.
Evidently, \eq \eqref{ALP_pot} is minimized for $a/f_a = \bar \theta$ \cite
{Vafa:1984xg}, and so the axion resolves the strong CP problem. Key for
achieving this is the breaking of the axionic shift symmetry by QCD effects.

It should be emphasized that we identify as QCD axion any particle that
receives its main contribution to the potential from the operator \eqref
{topologicalDensityAxion}, irrespectively of its UV-completion.

\paragraph*{Review of gravitational axion}

Although the concrete implementations \cite
{Mielke:2006zp,Mercuri:2009zi,Mercuri:2009zt,Lattanzi:2009mg,
Castillo-Felisola:2015ema,Karananas:2018nrj,Karananas:2024xja} of how to
obtain an axion from gravity \cite
{Mielke:2006zp,Mercuri:2009zi,Mercuri:2009zt,Lattanzi:2009mg} differ, the
essence of the idea is the following. The starting point is the fact that GR
exists in different formulations~(see \cite
{Heisenberg:2018vsk,BeltranJimenez:2019esp,Rigouzzo:2022yan} for overviews).
All of them are fully equivalent in the absence of matter and for a simple
choice of gravitational action, \eg consisting only of the Einstein-Hilbert
term. 

Most commonly used is the metric -- or second-order -- formulation, in which
spacetime is fully described by curvature. However, strong arguments speak in
favor of the EC formulation, which is endowed with torsion in addition to
curvature \cite{Cartan:1922,Cartan:1923,Cartan:1924,Cartan:1925}. In
particular, EC gravity can be derived from gauging the global Lorentz
group \cite{Utiyama:1956sy,Kibble:1961ba,Sciama:1962}, which puts the
gravitational interaction on the same footing as the other SM forces. Related
to the gauging is the fact that coupling fermions to GR naturally leads to
the EC formulation of GR \cite{Kibble:1961ba} (see also \cite
{Rigouzzo:2023sbb, Barker:2024dhb}). 

In EC gravity, the minimal kinetic term of a spinor $\Psi$ splits as
(see \cite{Karananas:2021zkl,Rigouzzo:2023sbb})
\be
\label{fermionKinetic}
\frac{i}{2}\bar{\Psi}\gamma^\mu \mathcal{D}_\mu \Psi + \text{h.c.} 
= \left(\frac{i}{2}\bar{\Psi}\gamma^\mu \mathcal{\mathring{D}}_\mu \Psi 
+ \text{h.c.}\right) -\frac{1}{8} J_A^\mu  \hat{T}_\mu \;,
\ee
where $\mathcal{D}_\mu$ is the fermionic covariant derivative, $\mathcal
{\mathring{D}}_\mu$ denotes its torsion-free, \ie Riemannian, counterpart,
and we introduced the axial fermionic current
\be 
\label{axial_current}
J_A^\mu = \bar{\Psi} \gamma^5  \gamma^\mu  \Psi \;.
\ee 
Moreover, $\gamma^\mu$ and $\gamma^5$ are gamma-matrices and $\hat{T}_
{\alpha}=\epsilon_{\alpha \beta \mu \nu}T^{\beta \mu \nu}$ corresponds to the
axial torsion vector, with $T^\mu_{ \ \alpha \nu}$ derived from the
connection $\Gamma^\mu_{ \ \alpha \nu}$ as  $T^\mu_{ \ \alpha \nu}\equiv 1/2
(\Gamma^\mu_{ \ \alpha \nu} - \Gamma^\mu_{ \  \nu \alpha})$; see
Appendix~\ref{app:conventions} for more details about conventions.
 
As is well-known, \cf \eg \cite
{Neville:1979rb,Sezgin:1979zf,Hecht:1996np,
Karananas:2014pxa,BeltranJimenez:2019hrm,Pradisi:2022nmh,Mikura:2023ruz,
Barker:2024dhb,Karananas:2024xja,Karananas:2025xcv,Barker:2025qmw}, axial
torsion contains a pseudoscalar degree of freedom $\phi$, which can be
heuristically extracted as $\hat{T}_{\alpha} = \partial_\alpha \phi$. As
evident from \eq \eqref{fermionKinetic}, $\phi$ couples to all fermions via
\be
\label{fundamentalOperator}
\mathcal{L} \supset  J_A^\mu \partial_\mu \phi  \;. 
\ee
Now we can integrate by parts and use the non-conservation of the chiral
current, ${\mathring{D}}_\mu J_A^\mu \supset \alpha/4\pi\, \text{Tr}\, G_
{\mu\nu} \tilde{G}^{\mu\nu}$, to conclude that the action contains a term
\be
\label{effectiveLIncomplete}
\mathcal{L} \supset \f{\alpha}{4\pi} \phi\, 
\text{Tr}\, G_{\mu\nu} \tilde{G}^{\mu\nu} \;.
\ee
When this is compared  with \eq \eqref{topologicalDensityAxion}, it appears
that $\phi$ couples to the topological density $\text{Tr}\, G_{\mu\nu} \tilde
{G}^{\mu\nu}$ in exactly the same way as the QCD axion (with the
identification $\phi= a/f_a$). Therefore, one is immediately tempted to
conclude that $\phi$ can assume the role of an axion, as was done in \cite
{Mielke:2006zp,Mercuri:2009zi,Mercuri:2009zt,Lattanzi:2009mg,
Castillo-Felisola:2015ema,Karananas:2018nrj,Karananas:2024xja}.

For completeness, we mention that this approach can only succeed if $\phi$
corresponds to a propagating degree of freedom, \ie if it possesses a kinetic
term.  This can be achieved  \eg with the addition of the term $\tilde
{R}^2$ \cite{Karananas:2024xja}, where $\tilde{R}$ is the pseudoscalar
(Holst) curvature defined in Appendix~\ref{app:EC_full_action}. Since this
point is inessential for our investigation, we shall not discuss it further.

\paragraph*{Proof to the contrary}

We will now show why the seemingly innocuous argument above does not go
through. To this end, it suffices to realize that before any manipulations,
the operator $ \partial_\mu \phi J_A^\mu$ as shown in \eq \eqref
{fundamentalOperator}, is \emph{manifestly} and \emph
{exactly} shift-symmetric. No effect, perturbative or non-perturbative, can
break this symmetry. 

Thus, the coupling~\eqref{fundamentalOperator} does not lead to a potential
for $\phi$;\footnote{The absence of the interaction~(\ref
{effectiveLIncomplete}) allows one to immediately match the Lagrangian to the
chiral one~\cite{Gasser:1984gg}. The would-be axion enters only through the
covariant derivative of the meson octet, meaning that no potential is
generated for it.}~and without a potential, there is no mechanism to
dynamically relax the CP violating angle to zero. Hence, the operator~\eqref
{fundamentalOperator} does not suffice to turn $\phi$ into an axion.

\paragraph*{What has gone wrong?} 

What remains to be done is demonstrate that the argument based on \eq \eqref
{effectiveLIncomplete} is incomplete. The key point is that there is another
contribution to the chiral anomaly from the mass $m_\Psi$ of the fermion
(see \eg \cite{Alvarez-Gaume:1983ihn,DiLuzio:2020wdo}):
\be
\label{eq:anomalous_divergence}
{\mathring{D}}_\mu 	J_A^\mu = \frac{\alpha}{4\pi} 
\text{Tr}\, G_{\mu\nu} \tilde{G}^{\mu\nu} 
+ 2 i  \bar {\Psi}m_\Psi  \gamma^5 \Psi \;.
\ee
Therefore, the term $\phi {\mathring{D}}_\mu  J_A^\mu$ entails two
contributions to the axionic potential --- one from the interaction with
gluons and the other from the Yukawa coupling to the fermions. Consistency
with the exact shift symmetry demands that they cancel exactly~\cite
{Kim:2008hd}. 

This argument can also be made by considering the~(anomalous) axion-dependent
chiral rotation
\begin{equation}
\label{fermion_rotation}
\Psi_{L,R} \mapsto e^{\mp i  \phi/2}\Psi_{L,R} \ ,
\end{equation}
where $\Psi_{L,R}= 1/2(1\mp \gamma_5) \Psi$. 
In this way,  one trades the derivative coupling $\partial_\mu \phi
J_A^\mu$ for the following new terms
\be
\label{new}
\mathcal{L}\supset - \left(e^{-i \phi }\bar\Psi_L m_\Psi \Psi_R  
+{\rm h.c.}\right)+\frac{\alpha}{4\pi} \, \phi\, 
\text{Tr}\, G_{\mu\nu} \tilde {G}^{\mu\nu} \;,
\ee
which when expanded around $\phi\approx 0$, give rise to exactly the same
contributions to the two-point function of $\phi$ as \eq (\ref
{eq:anomalous_divergence}).

A few remarks are in order at this point. First, it should not come as a
surprise that the fermion mass plays a crucial role. Indeed, the topological
term \eqref{topologicalDensity} does not lead to observable CP violation if
at least one massless fermion is present \cite
{tHooft:1976rip,tHooft:1976snw,Jackiw:1976pf,Callan:1976je,Peccei:1977hh}.
Second, one can wonder if an analogous argument can be made for the ``true''
axion coupling in \eq \eqref{topologicalDensityAxion}. Since $\text{Tr}\, G_
{\mu\nu} \tilde{G}^{\mu\nu} = \partial_\mu K^\mu$ with Chern-Simons current
$K^\mu$, one could try to integrate by parts and rewrite
\be
a \text{Tr}\, G_{\mu\nu} \tilde{G}^{\mu\nu} \propto 
-  K^\mu \partial_\mu a \;.
\ee
So one might be tempted to conclude that, also here, the axion obeys an exact
shift symmetry. However, the crucial difference with \eq \eqref
{fundamentalOperator} is that $K^\mu$ is not gauge-invariant, and so it is
impossible to define the fundamental Lagrangian respecting the local SU
(3) symmetry in terms of the Chern-Simons current.

\paragraph*{Quantum corrections} 

In fact, there is a loophole in the previous argument that can be exploited to
turn the gravitational scalar $\phi$ into the QCD axion. Inspired by the
actual way that the field couples to fermions in EC gravity
(see Appendix~\ref{app:EC_full_action}), let us generalize the
interaction \eqref{fundamentalOperator} as
\begin{align}
\label{eq:action_ALP_torsion_general}
\mathcal{L} &\supset \mathcal Z_V(\phi)\partial_\mu \phi  J^\mu_V 
+ \mathcal Z_A(\phi)\partial_\mu \phi  J^\mu_A  \nonumber\\ 
& +\mathcal Z_{VV}(\phi)\frac{{J^\mu_V}^2}{M_P^2}
+\mathcal Z_{AA}(\phi)\frac{{J^\mu_A}^2}{M_P^2}  
+\mathcal Z_{AV}(\phi)\frac{J_{A\,\mu} J^\mu_V }{M_P^2}\;,
\end{align} 
where $M_P$ is the Planck mass and $\mathcal Z(\phi)$ are a priori arbitrary
functions of the dimensionless $\phi$. The axial current $J_A^\mu$ was
defined previously in \eqref{axial_current}, while 
\begin{equation}
\label{vector_current}
J^\mu_V = \bar{\Psi} \gamma^\mu  \Psi
\end{equation} 
is its vectorial counterpart. The crucial point is that \eq \eqref
{eq:action_ALP_torsion_general} does not feature any shift symmetry of
$\phi$, and so the restriction of our previous argument is evaded.

\paragraph*{Perturbative effects} 

\begin{figure}
	\centering
	 \begin{minipage}{.49\linewidth}
	 	\centering
		\includegraphics[scale=.4]{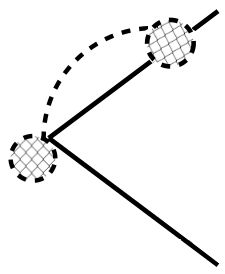}
	 \end{minipage}
	 \begin{minipage}{.49\linewidth}
		\centering
		\includegraphics[scale=.25]{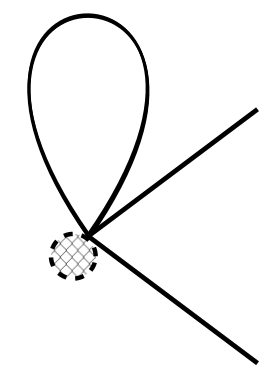}\hfill
	 \end{minipage}
	\caption{Exemplary Feynman diagrams generating Yukawa-type couplings
     between the $\phi$ and quarks. We used a blob to denote $\phi$
     insertions, which should be dressed appropriately depending on the
     $\mathcal Z$-functions that appear in the interactions. As usual, dashed
     and solid lines stand for scalar and fermionic fields, respectively. In
     order to obtain Yukawa couplings with $\gamma_5$, the corresponding
     diagrams should involve both the axial $J_A^\mu$ and vectorial $J_V^\mu$
     fermionic currents.} 
     \label{fig:feynman_diagrams}
\end{figure}

Since the factors $\mathcal Z$ in \eq \eqref
{eq:action_ALP_torsion_general} depend on $\phi$, these interactions induce
anomalous Yukawa-type couplings between the $\mathcal Z(\phi)$-functions and
quarks, of the schematic form 
\be
\label{eq:Induced_Yukawas_ALP}
\mathcal{L}_{\rm Yukawa} \propto 
i \,f(\phi) \bar\Psi_L m_\Psi \Psi_R + {\rm h.c.}\ ,
\ee
where $f(\phi)$ controls the strength and precise form of the interaction; it
is related to the various $\mathcal Z(\phi)$'s. Terms as in~(\ref
{eq:Induced_Yukawas_ALP}) appear already at the one-loop level, due to
diagrams in \fig \ref{fig:feynman_diagrams}. In this respect, what we
encounter here is close in spirit to the KSVZ construction~\cite
{Kim:1979if,Shifman:1979if}, where the axion couplings to the SM quarks are
generated at the loop level.  The left diagram in \fig
\ref{fig:feynman_diagrams} gives a contribution to $f(\phi) \propto  \mathcal Z_V
(\phi)  \mathcal Z_A(\phi)$, and the right graph $f(\phi) \propto  \mathcal
Z_{AV}(\phi)$.  The coefficients in front of the $\mathcal Z$-factors are
divergent and thus depend on the UV completion of the theory, which is
power-counting nonrenormalizable.

We therefore see that $\phi$ can indeed communicate nontrivially with QCD,
which is a necessary requirement for solving the strong CP problem by
gravity. The induced effective potential for $\phi$ and $\bar\theta$ is given
by \eq \eqref{ALP_pot} with $a/f_a\mapsto f(\phi)$, \ie
\be
\label{eq:gravi_ALP_pot}
V = -m_\pi^2 f_\pi^2 \sqrt{1-\frac{4 m_u m_d}{(m_u+m_d)^2}
\sin^2\left(\frac{1}{2}\left(\bar\theta - f(\phi)\right)\right)} \;.
\ee
If $V$ were the only contribution to the potential, its minimum at $\phi = f^
{-1}\left(\bar\theta\right)$ could lead to a dynamical vanishing of CP
violation.

\begin{figure}
    \centering
     \begin{minipage}{.49\linewidth}
        \centering
        \includegraphics[scale=.45]{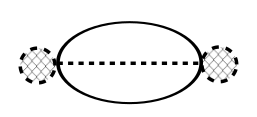}\hfill
     \end{minipage}
     \begin{minipage}{.49\linewidth}
        \centering
        \includegraphics[scale=.45]{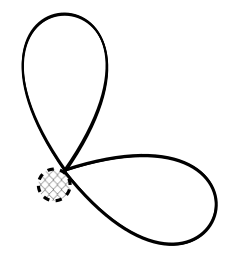}\hfill
     \end{minipage}
    \caption{Exemplary two-loop Feynman diagrams, originating from the
     couplings between the gravitational scalar and  fermions in  \eq \eqref
     {eq:action_ALP_torsion_general}, that feed into the potential for
     $\phi$.}
\label{fig:eff_pot_diagram}
\end{figure}

Several conditions must be satisfied for this solution to be viable. First,
the domain of possible values of the function $f(\phi)$ should include the
interval $[0,2\pi)$ to compensate for any value of the $\theta$ angle, \ie
it should be $f(\phi)\sim 1$. Second, the QCD contribution to the axion
potential should be dominant, \ie the gravitationally induced axion mass
must be much smaller than its QCD value. Besides a possible tree-level axion
mass, the interactions in \eq \eqref{eq:action_ALP_torsion_general} generate
the axion potential as well, see \fig \ref{fig:eff_pot_diagram}. They lead to
two-loop  contributions $\delta V \propto  \mathcal Z_V(\phi)^2,~   \mathcal
Z_A(\phi)^2, ~ \mathcal Z_{VV}(\phi)$ and $\mathcal Z_{AA}(\phi)$. Similarly
to the function $f(\phi)$, the coefficients in front of these factors depend
on the UV completion of the theory. To solve the strong CP problem, we must
have $\delta V \ll \Lambda_{\rm QCD}^4$, where $ \Lambda_{\rm QCD}\sim 100$
MeV is the QCD scale.

Let us consider different possibilities for the UV completion. First, let us
make a naive estimate of different divergent diagrams, assuming that they are
all saturated at some universal UV cutoff $\Lambda$, much exceeding the
electroweak scale, which can be as large as the Planck scale. Then, omitting
numerical coefficients $\mathcal O(1)$,
\be
\label{f}
f(\phi)\sim \f{\Lambda^2}{(4\pi f_a)^2}\left( \mathcal Z_V \mathcal Z_A
+\mathcal Z_{AV}\f{f_a^2}{M_P^2}\right)\,,
\ee
and
\be
\label{V}
\delta V\sim \f{m_\Psi^2 \Lambda^4}{(16\pi^2f_a)^2}\left((\mathcal Z_A^2
+\mathcal Z_V^2)+(\mathcal Z_{AA}+\mathcal Z_{VV})\f{f_a^2}{M_P^2}\right)\,.
\ee 
Clearly, the requirement $f(\phi)\sim 1$ leads to $\delta V\gg \Lambda_
{\rm QCD}^4$. So, the conditions formulated in the previous paragraph cannot
be met if the construction of the UV complete theory follows the naive power
counting in the estimates of the Feynman graphs, unless different
contributions to $\delta V$ are severely fine-tuned.

Second, let us assume that the complete theory is such that all power-like
divergences are irrelevant~\cite{Shaposhnikov:2007nj, Farina:2013mla}, which
can be implemented by using \eg dimensional regularization~(DimReg). Then, an
estimate of $f(\phi)$ and $\delta V$ is given by the formulas (\ref{f},\ref
{V}) with the replacement $\Lambda \to m_{\psi}$. So, any communication with
QCD due to the interactions~(\ref{eq:action_ALP_torsion_general}) is
practically nonexistent. In other words, if the only perturbative
contributions were associated with the interactions (\ref
{eq:action_ALP_torsion_general}), no solution to the strong CP problem would
be achieved: roughly speaking, the nearly massless gravitational scalar field
in this case is ``orthogonal'' to the required axion direction in the field
space.

\paragraph*{Weyl-invariant $R^2$ gravity}

\begin{figure}
\centering
\includegraphics[scale=.45]{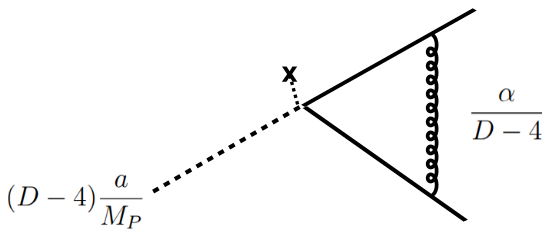}
\caption{Contribution to the axion-fermion coupling from evanescent operators.
 The factor $(D-4)$ multiplying $a/M_P$ comes from the expansion of $\left
 (M_P^2+M_Pa\right)^{D-4}$ around $D=4$ and $a=0$. This is compensated by the
 divergent loop factor, leading to a finite contribution. As before, dashed
 and solid lines stand for scalar and fermionic fields, respectively; the
 wavy line denotes gluons, while $\pmb{\times}$~is the vacuum expectation
 value of the Higgs.}
\label{fig:evanescent}
\end{figure}

Now we turn to the Weyl-invariant EC gravity as proposed in \cite
{Karananas:2024xja}, and briefly presented across Appendices~\ref
{app:EC_full_action} and~\ref
{app:weyl_invariant_einstein_cartan_gravity}. The self-consistency of the
theory requires that quantum corrections be treated in a manner that
preserves the Weyl symmetry. This singles out the use of the
``scale-invariant'' prescription~\cite{Englert:1976ep,Shaposhnikov:2008xi} to
regularize Feynman diagrams, where divergences are removed in a
scale-invariant way. The simplest realization is based on DimReg. The theory
is extended to $D \neq 4$ dimensions in a Weyl-invariant way, by
multiplication of the terms in the classical Lagrangian with appropriate
functions of the scalar fields present (Higgs $h$ taken in the unitary gauge,
dilaton/spurion $\chi$ that generates the Planck scale $M_P$, and axion
$a$).\footnote{In the standard DimReg, the dimensional 't Hooft parameter
$\mu$ --- the renormalization scale --- breaks the Weyl symmetry explicitly.}
For example, the Weyl-invariant Higgs-fermion interaction can be written
as \cite{Shaposhnikov:2025}
\be
\int \diff^D x \sqrt{g} \bar{\Psi}\Psi h \chi^{D-4} F(a/\chi,h/\chi)\,,
\ee 
where $F\to1$ when $D \to 4$, with a similar construction for the rest of the
terms in the classical action. The finite radiative corrections are derived
as in the 't Hooft minimal subtraction scheme by removing the poles in $D-4$.
The results depend on the function $F$, which determines the UV completion of
the theory. It cannot be fixed by a symmetry principle. What is important is
that the effective action (in the Einstein frame) contains the
higher-dimensional operators induced by these evanescent operators: the
factors of $D-4$ in the interaction strength are compensated by the poles in
$D-4$ appearing because of the loops \cite{Shaposhnikov:2009nk}. As a result,
the first and second terms which appear in eq.~(\ref{new}) are generated, but
now with uncorrelated numerical coefficients, depending on the function $F$.
To illustrate this point, in \fig~\ref{fig:evanescent} we have depicted how a
Yukawa-type mixing can be generated at one loop. Consequently, the required
coupling of the gravitational axion to these operators is already present in
the Weyl-invariant EC theory, enabling the gravitational solution of the
strong CP problem.

The discussion above was purely perturbative. The gravitational axion will
remain a viable option only if non-perturbative gravity contributions to the
axionic potential remain negligible. Conversely, it is conceivable that a
non-trivial coupling of the gravitational axion to QCD is generated by
non-perturbative phenomena, \eg via gravi-scalar instantons
\cite{Shaposhnikov:2018xkv,Shaposhnikov:2018jag,Shaposhnikov:2020geh}
 (see also~\cite{Karananas:2020qkp}).

\paragraph*{Outlook}

A particularly appealing candidate for the QCD axion is a scalar field arising
purely from gravity. Such realizations are minimal in that the field is
already present in the gravitational sector and automatically couples to
fermionic currents, and thus, naively, also to QCD via the chiral anomaly.
This feature distinguishes them from more conventional axion models that rely
on extra symmetries or fields.

We have shown, however, that previous arguments for how a QCD axion can be
generated from gravity are ultimately flawed. Furthermore, addressing the
shortcomings of these proposals appears to be significantly more challenging
than one might have anticipated.

Concretely, our findings can be summarized as follows:

$\bullet$~In theories where the would-be axion couples only derivatively to
   fermionic currents, its shift symmetry remains unbroken and so its
   presence is irrelevant for the strong CP problem. 

$\bullet$~If the coupling to fermions explicitly breaks shift symmetry, then
   quantum corrections can, in principle, open the communication with QCD.
   Generically, however, interactions are either too weak to be viable or 
   suffer from a severe fine-tuning problem. 

$\bullet$ These results leave one  avenue for realizing a viable
  gravitational axion in the framework of the Weyl-invariant Einstein-Cartan
  gravity~\cite{Karananas:2024xja}, where the coupling to QCD can arise from
  quantum Weyl invariance.

\begin{acknowledgments}

\paragraph*{Acknowledgments} 

The work of S.Z.~was supported by the European Research Council Gravites
Horizon Grant AO number: 850 173-6.

\textbf{Disclaimer:} Funded by the European Union. Views and opinions
 expressed are however those of the authors only and do not necessarily
 reflect those of the European Union or European Research Council. Neither
 the European Union nor the granting authority can be held responsible for
 them.
\end{acknowledgments}

\appendix
\section{Conventions} 
\label{app:conventions}

Our conventions are as in \cite{Rigouzzo:2023sbb,Karananas:2024xja}. The
metric $g_{\mu\nu}$ has signature $(-1,+1,+1,+1)$, and for the gamma matrices
we use
\begin{align} 
\label{gammaConvention}
&\left\{\gamma_\mu, \gamma_\nu \right\} = - 2 g_{\mu\nu} \;,\\
&\gamma_5 = -i \gamma^0 \gamma^1 \gamma^2 \gamma^3 = 
i \gamma_0 \gamma_1 \gamma_2 \gamma_3 \;,
\end{align}   
while the Levi-Civita symbol is
\be
\label{epsilonConvention}
\tilde{\epsilon}_{0123}=1=-\tilde{\epsilon}^{0123} \;,
\ee 
and we define the density 
\be
\epsilon^{\m\n\rho\s} = 
\f{\tilde{\epsilon}^{\m\n\rho\s}}{\sqrt{g}} \ ,
\ee
with $g=-{\rm det}(g_{\m\n})$. The curvature tensor reads
\be
R^{\rho}_{~\s\m\n} = \p_{\m}\G^{\rho}_{~\n\s}-\p_{\n}\G^{\rho}_{~\m\s} 
+ \G^{\rho}_{~\m\lambda}\G^{\lambda}_{~\n\s} 
- \G^{\rho}_{~\n\lambda}\G^{\lambda}_{~\m\s} \ ,
\ee
where $\G^\m_{~\n\rho}$ is the (metric-compatible) connection which is not
symmetric under the exchange of the lower indexes. The scalar and
pseudoscalar curvatures $R$ and $\tilde R$  are defined as
\begin{align}
\label{eq:app_curvature_scalars_defs}
&R = g^{\s\n}\delta^{\m}_{\rho} R^{\rho}_{~\s\m\n} \ ,
~~~\tilde R= \epsilon^{\rho\s\m\n} R_{\rho\s\m\n} \ . 
\end{align}
Finally, we work in natural units $\hbar=c=1$, but keep explicit the Planck
mass $M_P=1/\sqrt{8\pi G}$ with Newton's constant $G$.

\begin{widetext}

\onecolumngrid

\section{Full action of quadratic Einstein-Cartan gravity}
\label{app:EC_full_action} 

The most general Lagrangian in EC gravity that is at most quadratic in
derivatives and propagates a massless graviton and a pseudoscalar spin-0 mode
of gravitational origin reads\,\footnote{See~\cite{Barker:2025qmw} for the
extended action propagating an additional scalar.}~(we do not include a
cosmological constant term, which is inconsequential for the following)
\begin{align}
\label{eq:app_grav_lagr}
M_P^{-2}\mathcal L_{\rm gr.} = \f{R}{2} &+ Q  \tilde R 
+ \f{M_P^{-2}}{\tilde f^2}\tilde R^2 \nonumber\\ 
&+\f{C_{TT}}{3}T_\m T^\m -\f{C_{\hat T\hat T}}{48}\hat T_\m \hat T^\m 
-\f{C_{\tau\tau}}{4}\tau_{\m\n\rho}\tau^{\m\n\rho} 
+\f{2C_{T\hat T}}{3}\hat T_\m T^\m 
+ \f{\tilde C_{\tau\tau}}{2}\epsilon^{\m\n\rho\s}
\tau_{\lambda\m\n}\tau^\lambda_{~\rho\s}  \ ,
\end{align} 
with $Q, \tilde f, C_{TT}, C_{\hat T \hat T}, C_{T\hat T}, C_
{\tau\tau}, \tilde C_{\tau\tau}$ arbitrary dimensionless constants. The
torsion vector $T$, pseudovector $\hat T$ and reduced tensor
$\tau$\,\footnote{The reduced torsion tensor satisfies
\be
\tau_{\m\n\rho} = -\tau_{\m\rho\n} \ ,~~~\tau^\m_{~\m\n}= 0 \ ,
~~~\epsilon^{\m\n\rho\s}\tau_{\n\rho\s} = 0 \ .
\ee}
appearing in the above are defined as
\begin{align}
&T^\m = T^{\n\m}_{~~\n} \ ,
~~~\hat T^\m= \epsilon^{\m\n\rho\s} T_{\n\rho\s} \ ,
~~~\tau_{\m\n\rho} = \f 2 3
\left(T_{\m\n\rho} - T_{[\n\rho]\m} -T_{[\n}g_{\rho]\m}\right)\ ,
\end{align}
where 
\be
T^\m_{~\n\rho} =\f 1 2(\G^\m_{~\n\rho}-\G^\m_{~\rho\n}) \ ,
\ee
is the torsion tensor; square brackets stand for antisymmetrization of the
enclosed indexes.

It is convenient to decompose the curvature scalars \eqref
{eq:app_curvature_scalars_defs} into Riemannian (torsion-free) plus
post-Riemannian (torsionful) pieces
\begin{align}
\label{eq:app_resolved_curvatures}
&R = \mathring R + 2\mathring \nabla_\m T^\m -\f 2 3 T_\m T^\m 
+\f{1}{24} \hat T_\m \hat T^\m +\f 1 2 \tau_{\m\n\rho}\tau^{\m\n\rho} \ ,
~~~\tilde R = -\mathring \nabla_\m \hat T^\m +\f 2 3 \hat T_\m T^\m 
+ \f 1 2 \epsilon^{\m\n\rho\s}\tau_{\lambda\m\n}\tau^{\lambda}_{~\rho\s} \ ,    
\end{align}
where $\mathring R$ is the Riemannian Ricci scalar and $\mathring \nabla_\m$
denotes the torsion-free covariant derivative.

To bring~(\ref{eq:app_grav_lagr}) into its metric-equivalent form~\cite
{Karananas:2021zkl,Karananas:2021gco,Karananas:2024xja,Karananas:2025xcv,
Barker:2025qmw}, we introduce a dimensionless auxiliary field $\vp$ to
rewrite it as 
\begin{align}
\label{eq:app_grav_lagr_auxiliary}
M_P^{-2}\mathcal L_{\rm gr.} =\f{R}{2} &+ \left(Q +\vp\right)  \tilde R 
-\f{\tilde f^2 M_P^2\vp^2}{4} \nonumber\\
&+\f{C_{TT}}{3}T_\m T^\m -\f{C_{\hat T \hat T}}{48}\hat T_\m \hat T^\m  
-\f{C_{\tau\tau}}{4}\tau_{\m\n\rho}\tau^{\m\n\rho}
+\f{2C_{T\hat T}}{3}\hat T_\m T^\m 
+ \f{\widetilde C_{\tau\tau}}{2}
\epsilon^{\m\n\rho\s}\tau_{\lambda\m\n}\tau^\lambda_{~\rho\s}  \ ,
\end{align} 
and then plug in Eqs.~(\ref{eq:app_resolved_curvatures}), to end up with
\begin{align}
\label{eq:app_grav_lagr_final}
M_P^{-2}\mathcal L_{\rm gr.} =\f{\mathring R}{2} &
- \vp \mathring\nabla_\m \hat T^\m -\f{\tilde f^2 M_P^2\vp^2}{4} 
+\mathring\nabla_\m\left(T^\m -Q\hat T^\m\right) \nonumber\\
&+\f{c_{TT}}{3}T_\m T^\m-\f{c_{\hat T\hat T}}{48}\hat T_\m \hat T^\m
-\f{c_{\tau\tau}}{4}\tau_{\m\n\rho}\tau^{\m\n\rho}
+\f{2(q+\vp)}{3}\hat T_\m T^\m+\f{\tilde c_{\tau\tau}+\vp}{2}
\epsilon^{\m\n\rho\s}\tau_{\lambda\m\n}\tau^\lambda_{~\rho\s} \ ,
\end{align}
where 
\be
c_{ij} = C_{ij} - 1 \ ,
~~~\tilde c_{\tau\tau}=\tilde C_{\tau\tau}+Q\ ,
~~~q = C_{T\hat T} + Q \ .
\ee
 
The above should be supplemented with the following fermionic Lagrangian~\cite
{Karananas:2021zkl}
\begin{align}
\label{eq:app_fermion_Lagr}
\mathcal L_{\Psi} = \f i 2 \Big(\bar \Psi_L \gamma^\m \mathring D_\m \Psi_L &
+\bar \Psi_R \gamma^\m \mathring D_\m \Psi_R +2i \bar\Psi_L m_{\Psi}\Psi_R\Big)  
+{\rm h.c.} \nonumber\\
&+\left(\tilde\z_V J^\m_V +\tilde\z_A J^\m_A\right)T_\m + \left(\z_V J^\m_V 
+\z_A J^\m_A\right)\hat T_\m    \ ,
\end{align} 
where the covariant derivative $\mathring D_\m$, in addition to the SU
(3) gauge field, contains also the torsion-free spin connection. We denoted
with $\z_A=z_A-1/8$, and $\z_{V}$, $z_A$ and $\tilde\z_{A,V}$ are nonminimal
fermionic couplings to torsion. 

The total Lagrangian of the system under consideration therefore reads
\be
\label{eq:app_full_Lagr}
\mathcal L = \mathcal L_{\rm gr.} + \mathcal L_\psi \ ,
\ee 
with $\mathcal L_{\rm gr.}$ and $\mathcal L_\psi$ given by~(\ref
{eq:app_grav_lagr_final}) and~(\ref{eq:app_fermion_Lagr}), respectively.
Since torsion enters algebraically, it can be integrated out via
the equations of motion for $T,\hat T,\tau$. This gives
\begin{align}
\mathcal L = \f{M_P^2}{2}\mathring R &
+\f{12M_P^2 c_{TT}}{c_{TT}c_{\hat T\hat T}+16(q+\vp)^2}(\p_\m\vp)^2 
- \f{\tilde f^2 M_P^4}{4}\vp^2 \nonumber \\ 
&+\f i 2 \left(\bar \Psi_L \gamma^\m \mathring D_\m \Psi_L 
+\bar \Psi_R \gamma^\m \mathring D_\m \Psi_R 
+2i \bar\Psi_L m_{\Psi}\Psi_R\right)  +{\rm h.c.}  \nonumber \\ 
& + \mc Z_V(\vp)\p_\m \vp J^\m_V + \mc Z_A(\vp)\p_\m \vp J^\m_A 
+\mc Z_{VV}(\vp)\f{{J^\m_V}^2}{M_P^2}+ \mc Z_{AA}(\vp)\f{{J^\m_A}^2}{M_P^2} 
+ \mc Z_{AV}(\vp)\f{J_{A\,\m} J^\m_V }{M_P^2} \ ,
\label{app_equivalent_theory} 
\end{align}
where
\begin{align}
\label{eq:app_ZV}
&\mc Z_{V,A}(\vp) = \f{24\left(c_{TT}\z_{V,A}
-(q+\vp)\tilde\z_{V,A}\right)}{c_{TT}c_{\hat T\hat T}+16(q+\vp)^2} \ ,\\
&\mc Z_{VV,AA}(\vp) = 
\f{12 c_{TT}\z_{V,A}^2-\f{3c_{\hat T\hat T}}{4}\tilde\z_{V,A}^2
-24\z_{V,A}\tilde\z_{V,A}(q+\vp)}{c_{TT}c_{\hat T\hat T}+16(q+\vp)^2} \ ,\\
&\mc Z_{AV}(\vp) = \f{24c_{TT}\z_V\z_A-\f{3c_{\hat T\hat T}}{2}
\tilde\z_V\tilde\z_A-24(\z_V\tilde\z_A+\tilde\z_V\z_A)(q+\vp)}
{c_{TT}c_{\hat T\hat T}+16(q+\vp)^2} \ .
\label{eq:app_ZAV}
\end{align}

The nonlinearities of the kinetic term for $\vp$ can be removed by
introducing 
\be
\phi = \sqrt{-\f{3c_{TT}}{2}}
\log\left[\sqrt{c_{TT}c_{\hat T\hat T}+16(q+\vp)^2}-4(q+\vp) \right] \ , 
\ee 
meaning that the relevant parts of the Lagrangian in terms of the above boil
down to
\begin{align}
\label{eq:app_ALP_Lagr}
\mathcal L &\supset -\f{M_P^2}{2}(\p_\m \phi)^2 -V(\phi) 
+ \mc Z_V(\phi)\p_\m \phi J^\m_V + \mc Z_A(\phi)\p_\m \phi J^\m_A 
+\mc Z_{VV}(\phi)\f{{J^\m_V}^2}{M_P^2}+ \mc Z_{AA}(\phi)\f{{J^\m_A}^2}{M_P^2} 
+ \mc Z_{AV}(\phi)\f{J_{A\,\m} J^\m_V }{M_P^2} \ ,
\end{align}
with 
\be
V(\phi) = \f{\tilde f^2 M_P^4}{256}
\left(8q + e^{\sqrt{-\f{2}{3c_{TT}}}\phi} 
-c_{TT}c_{\hat T\hat T}e^{-\sqrt{-\f{2}{3c_{TT}}}\phi}\right)^2 \ ,
\ee 
and now
\begin{align}
&\mc Z_{V,A}(\phi) = \sqrt{-\f{3}{2c_{TT}}}\left(\f{2\left(\tilde\z_{V,A}
+4c_{TT}\z_{V,A}e^{\sqrt{-\f{2}{3c_{TT}}}\phi}\right)}
{1+c_{TT}c_{\hat T\hat T}e^{2\sqrt{-\f{2}{3c_{TT}}}\phi}} 
-\tilde\z_{V,A}\right) \ ,\\
&\mc Z_{VV,AA}(\phi) = \f{3e^{\sqrt{-\f{2}{3c_{TT}}}\phi}\left(4\z_{V,A}
-c_{\hat T\hat T}e^{\sqrt{-\f{2}{3c_{TT}}}\phi}\tilde\z_{V,A}\right)
\left(4c_{TT}\z_{V,A}e^{\sqrt{-\f{2}{3c_{TT}}}\phi}+\tilde\z_{V,A}\right)}
{\left(1+c_{TT}c_{\hat T\hat T}e^{2\sqrt{-\f{2}{3c_{TT}}}\phi}\right)^2} \ ,\\
&\mc Z_{AV}(\phi) = \f{6e^{\sqrt{-\f{2}{3c_{TT}}}\phi}
\left(\left(16c_{TT}\z_V\z_A-c_{\hat T\hat T}\tilde\z_V\tilde\z_A\right)
e^{\sqrt{-\f{2}{3c_{TT}}}\phi}+2(\z_V\tilde\z_A+\tilde\z_V\z_A)
\left(1-c_{TT}c_{\hat T\hat T}e^{2\sqrt{-\f{2}{3c_{TT}}}\phi}\right)\right)}
{\left(1+c_{TT}c_{\hat T\hat T}e^{2\sqrt{-\f{2}{3c_{TT}}}\phi}\right)^2} \ ,
\label{eq:ZAVa}
\end{align} 
following immediately from~(\ref{eq:app_ZV})-(\ref{eq:app_ZAV}). In an abuse
of notation we retain the same symbol $\mc Z$ for the coupling functions. It
is obvious from~(\ref{eq:app_ALP_Lagr}) that the canonical axion $a$ is
related to $\phi$ as
\be
a = M_P \phi \ .
\ee

\section{Weyl-invariant Einstein-Cartan gravity}
\label{app:weyl_invariant_einstein_cartan_gravity}

Of particular interest is the Weyl-invariant setting we introduced and studied
in details in~\cite{Karananas:2024xja}. In the broken phase,$^{\ref
{foot:unbroken_Weyl}}$ it corresponds to choosing in \eq \eqref
{eq:app_grav_lagr}
\be
\label{eq:app_Weyl_coefs_choice}
Q= C_{TT}=C_{\hat T\hat T}=C_{T\hat T}= 0 \;,
\ee
and in \eq \eqref{eq:app_fermion_Lagr}
\be
\tilde\z_V = \tilde\z_A=0 
\ee 

So the ALP sector is simply obtained from \eqs~(\ref{eq:app_ALP_Lagr}) to
(\ref{eq:ZAVa}) with these replacements, where we note that $c_{TT}=c_{\hat T\hat
T}=-1$. This results in the following potential for $\phi$
\be
\label{eq:app_Weyl_pot}
V(\phi) = \f{\tilde f^2 M_P^4}{64}\sinh^2\left(\sqrt{\f{2}{3}}\phi\right) \ ,
\ee 
while the coupling functions are given by 
\begin{align}
&\mc Z_{VV,AA}(\phi) =
- \f{12\z_{V,A}^2}{\cosh^2\left(\sqrt{\f 2 3} \phi\right)}\ ,
~~~\mc Z_{AV}(\phi) =
-\f{24\z_A\z_V}{\cosh^2\left(\sqrt{\f 2 3} \phi\right)}\ ,
~~~\mc Z_{V,A}(\phi) = 
-\f{2\sqrt 6 \z_{V,A}}{\cosh\left(\sqrt{\f 2 3} \phi\right)} \ .
\label{ZVa}
\end{align}
\end{widetext}

\footnotetext[5]{\label{foot:unbroken_Weyl}Before spontaneous Weyl symmetry
 breaking, the Lagrangian only contains terms quadratic in the scalar and
 pseudoscalar curvatures and reads
\be
\label{eq:app_Weyl_EC_Lagr}
\mathcal L_{\rm W} = \f{1}{f^2}R^2+ \f{1}{\tilde f^2}\tilde R^2 \ ,
\ee 
where for simplicity we omitted the cross-term $R\tilde R$; its inclusion
leads to a shifted potential for the axion~\cite
{Gialamas:2024iyu,Gialamas:2024iyu,Karananas:2025xcv}, and is irrelevant for
our discussions here. The $R^2$-term can be brought into a first-order form
in terms of an auxiliary dilaton/spurion $\chi$ as $R^2 \to \chi^2 R -
f^2\chi^4/4$. Weyl invariance allows to set $\chi=M_P/\sqrt{2}$~\cite
{Karananas:2024xja}, and~(\ref{eq:app_Weyl_EC_Lagr}) exactly reproduces~(\ref
{eq:app_grav_lagr}) for the choices~(\ref{eq:app_Weyl_coefs_choice}). }

\bibliography{Refs.bib}

\end{document}